\documentclass[preprint,authoryear,12pt]{elsarticle}

\usepackage{epsfig}

\usepackage{amssymb}
\usepackage{wasysym}

\usepackage[ps2pdf,%
a4paper=true,%
breaklinks=true,%
colorlinks=true,%
pdfauthor={S. Dasso and H. Asorey},%
pdftitle={The scaler mode in the Pierre Auger Observatory to study heliospheric modulation of cosmic rays}%
]{hyperref}

\journal{Advances in Space Research}

\begin{document}

\begin{frontmatter}



\title{The scaler mode in the Pierre Auger Observatory to study heliospheric modulation of cosmic rays}


\author{S. Dasso\fnref{d1}}
\address{Departamento de F\'{i}sica (FCEN-UBA) and Instituto de Astronom\'{i}a y F\'{i}sica del Espacio (FCEN-UBA,CONICET), Buenos Aires, Argentina.}
\ead{dasso@df.uba.ar}

\author{H. Asorey\fnref{d2}}
\address{Centro At\'{o}mico Bariloche (CNEA), U. N. de Cuyo and U. N. de R\'{i}o Negro, Bariloche, R\'{i}o Negro, Argentina.}
\ead{asoreyh@cab.cnea.gov.ar}

\author{for the Pierre Auger Collaboration\fnref{d3}}
\address{Observatorio Pierre Auger, Av. San Mart\'in Norte 304, 5613
Malarg\"ue, Argentina.}
\ead{auger\_spokesperson@fnal.gov}

\fntext[d1]{Tel. +54-11-4781-6755 ext 128, Fax: +54-11-4786-8114}
\fntext[d2]{Tel. +54-2944-44-5151 ext 38, Fax: +54-2944-44-5126}
\fntext[d3]{Full author list: http://www.auger.org/archive/authors\_2011\_05.html}

\begin{abstract}
The impact of the solar activity on the heliosphere has a strong influence on
the modulation of the flux of low energy galactic cosmic rays arriving at
Earth.  Different instruments, such as neutron monitors or muon detectors, have
been recording the variability of the cosmic ray flux at ground level for
several decades. Although the Pierre Auger Observatory was designed to observe
cosmic rays at the highest energies, it also records the count rates of low
energy secondary particles (the scaler mode) for the self-calibration of its
surface detector array.  From observations using the scaler mode at the Pierre Auger
Observatory, modulation of galactic cosmic rays due to solar transient activity
has been observed (e.g., Forbush decreases).  Due to the high total count rate
coming from the combined area of its detectors, the Pierre Auger Observatory (its
detectors have a total area greater than $16\,000$\,m$^2$) detects a flux of
secondary particles of the order of $\sim 10^8$\,counts per minute. Time
variations of the cosmic ray flux related to the activity of the heliosphere
can be determined with high accuracy.  In this paper we briefly describe the
scaler mode and analyze a Forbush decrease together with the interplanetary
coronal mass ejection that originated it.  The Auger scaler data are now
publicly available.
\end{abstract}

\begin{keyword}
Cosmic Rays \sep Cherenkov detectors \sep Particle detectors \sep Interplanetary Coronal Mass Ejections
\end{keyword}

\end{frontmatter}

\parindent=0.5 cm

\section{Introduction}
\label{SCintro}

Transport of galactic cosmic rays (CRs) in the heliosphere is one of the topics
of major interest in space physics and presents several unsolved questions.
Transport of CRs is modulated by different physical mechanisms, which can be
divided into large scale processes (related to the large scale heliospheric
magnetic field) and transient phenomena, such as those produced by transient
solar ejecta or interplanetary shocks.

As the flux of CRs decreases for higher energies, a large collecting surface is
required when detecting high energy CRs, so ground-based instruments have to be
used to study these elusive particles. For several decades, neutron monitors
and muon detectors have been the detectors of secondary particles commonly used
to determine fluxes of primary CRs with energies larger than $\sim 1$\,GeV.

The time variability of the cosmic ray flux has been systematically recorded
since the 1950s by neutron monitors. They have been crucial for understanding
different mechanisms of solar modulation of cosmic rays, such as the
anti-correlation between the sunspot number and cosmic rays intensity
\citep{Meyer_Simpson_1955}, the 27 days recurring intensity variations
associated with the impact of the solar rotation combined with coronal holes on
the interplanetary magnetic field \citep[see e.g.,][]{Simpson1998}, and Forbush
decreases (Fd) \citep{Forbush37}. Forbush decreases are significant depressions
observed in CR flux at Earth, which are generally observed in association with
the arrival to the Earth's environment of interplanetary shocks driven by huge
transient magnetic structures of solar origin, the so called Interplanetary
Coronal Mass Ejections (ICMEs) \citep[e.g.,][]{Cane00}.  Forbush decreases can
be also produced by ICMEs without shocks, and the structure of the time profile
of CR intensity is significantly different in cases with the presence of a
shock when compared to cases of ICMEs without a driven shock wave
\citep[e.g.,][]{wibberenz98}.  Despite several properties of the structure of a
Fd being relatively well understood, their recovery times are still not well
known \citep[e.g., ][]{Usoskin08}.

The heliosphere presents a variety of dynamical structures, which are not yet
satisfactorily understood.  In the last decades solar wind observations made
from spacecraft have helped to significantly improve our knowledge of these
structures.  Remote observations of the solar wind have serious limitations to
competently determine spatial magnetic distributions in the heliosphere.
Instead, 'in situ' solar wind observations can provide direct magnetic field
observations, but they have some difficulties in measuring global magnetic
structures because they can only observe local quantities (one point - multi
times) along the linear (one dimensional) trajectory of the probe in the solar
wind, and then they can combine spatial shapes and time evolution during the
observation period.

Alternatively, ground observations of cosmic rays of low energies can provide
precious information to complement 'in situ' information about the
interplanetary magnetic field. The combination of observations from both cosmic
ray ground observations and solar wind observations from space provides a good
opportunity to make detailed studies of interplanetary structures and their
effects on propagation of particles. 

The $1\,660 \times 10$\,m$^2$ detectors of the Pierre Auger Observatory measure
the flux of low energy cosmic rays arriving at Earth with a huge total
collecting area, more than $16\,000$\,m$^2$.  This provides a cosmic ray
detector of high accuracy, recording of the order of $\sim 10^8$\,counts per
minute, with a consequent very high statistical significance.

In Section \ref{SCpao} we briefly describe the Pierre Auger Observatory.
Section \ref{SCscalers} presents its scaler mode for observing low energy
particle fluxes.  A comparison of Auger scalers and neutron monitor
observations is given in Section \ref{SCcomparison}.  In Section \ref{SCicme}
we present an analysis of the interplanetary perturbation producing a Forbush
decrease observed at Auger.  Finally, in Section \ref{SCconclusion} we present
a summary and the conclusions of the paper.

\section{The Pierre Auger Observatory}\label{SCpao}
\label{pao}

\begin{figure}
\includegraphics*[width=12cm,angle=0]{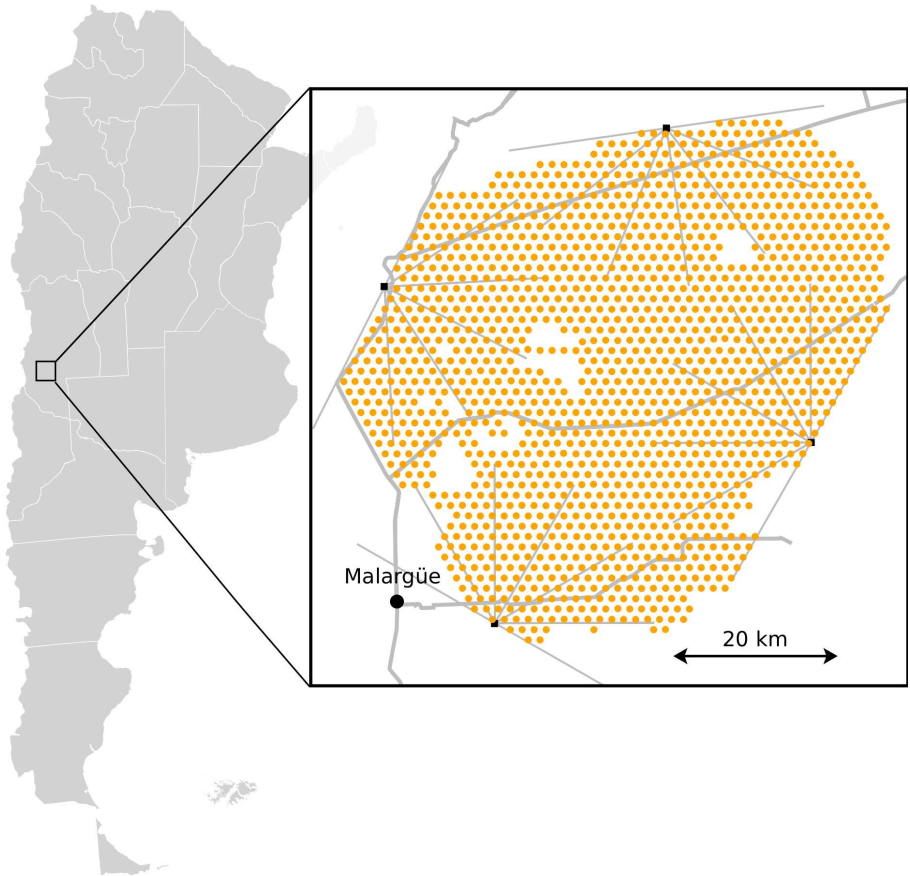}
\caption{Geographical location of the SD array of the Pierre Auger Observatory,
Malarg\"ue, Argentina.  The position of the four fluorescence telescope
buildings surrounding the SD array are indicated by the black squares, with
gray lines indicating the field of view of the six telescopes at each building.
The positions of $1\,660$ water-Cherenkov detectors (in the year 2009) are
marked with small dots.}
\label{figure_mapa_pao}
\end{figure}

The Pierre Auger Observatory \citep{ThePierreAugerCollaboration2004} was
designed to study the physics of cosmic rays at the highest energies.  It is
located in the west of Argentina (Malarg\"ue, $69.3^\mathrm{o}$\,W,
$35.3^\mathrm{o}$\,S, $1\,400$\,m a.s.l.) and it combines two techniques: (a)
the observation of the fluorescence light produced by secondary particles as
they propagate through the atmosphere, and (b) the direct measurement of
particles reaching ground level.  The layout of the Pierre Auger Observatory is
shown in Figure \ref{figure_mapa_pao}.

The interaction of a high-energy particle with the atmosphere produces a shower
formed by a huge number of secondary particles, called an Extensive Atmospheric
Shower (EAS).  The development of the EAS can be tracked by the Fluorescence
Detector \citep{ThePierreAugerCollaboration2010}, which consists of 24
telescopes grouped in modules of six telescopes at four different locations
(Los Leones, Coihueco, Loma Amarilla and Los Morados).

The ground level observations are based on measurements coming from an array of
Surface Detectors (SD), which covers a surface of $3\,000$\,km$^2$ where
$1\,660$ water-Cherenkov detectors are placed in a triangular grid with a
spacing of $1\,500$\,m.  Each water-Cherenkov detector consists of a
polyethylene tank ($10$\,m$^2$ area) containing $12$\,m$^3$ of high-purity
water in a highly-reflective Tyvek$\textsuperscript{\textregistered}$ liner bag
\citep{Allekotte08}.

Cherenkov radiation produced by the charged particles passing through the water
volume in each detector is measured by three photomultiplier tubes (PMTs) and
the signals from the PMTs are processed with a sampling rate of $40$\,MHz by
six 10-bit flash analog-to-digital converters (FADC), and sent by a radio link
to the central data acquisition system (CDAS) in Malarg\"ue city, Argentina.  A
GPS system is used for timing and synchronization. 

Particles interact with the water producing pulses of different size (with the
area of the pulse related to the energy deposited by the particle in the
detector), which are recorded to construct histograms.  Scalers record the
total count of signals above a low threshold (see next section).  More details
about SD detectors can be found in \cite{Allekotte08} and about scalers in
\cite{jinst}.

\section{Pierre Auger Observatory and Galactic Cosmic Rays Counting}\label{SCscalers}

In March 2005, a ``single particle technique'' mode was implemented for the
full array of SD detectors at the Pierre Auger Observatory. It consists in
recording the rate of signals associated with the energy deposited by secondary
particles, the scaler mode \citep{jinst}, and it measures low energy radiation
which is mainly useful for monitoring the long-term stability of the detector,
for searching transient events (such as gamma ray bursts or Forbush decreases),
and for studying long-term trends in the heliospheric modulation of cosmic rays
during the solar cycle.

Two different scaler modes have been implemented at Auger.  In the first one
(period I: from March 1, 2005 to September 20, 2005) it counted the total
number of signals per second in each detector above a threshold that
corresponds to secondary particles with deposited energies ($E_d$) larger than
$\sim 15$\,MeV.  In Period II (from September 30, 2005 to present) an upper
bound was introduced to reduce the influence of muons, and the deposited
energies considered correspond to 15\,MeV $\lesssim E_d \lesssim$ 100\,MeV (see
Figure 1 of \cite{icrc_china}).  The full SD array was completed in 2008, with
a collecting area of more than $16\,000$\,m$^2$ and a scaler counting rate of
$\sim 2 \times 10^8$\,counts\,min$^{-1}$.  More details about scaler modes can
be found in \citet{Asorey2009,jinst,icrc_china}.

\begin{figure}
\begin{center}
\includegraphics*[width=14cm,angle=0]{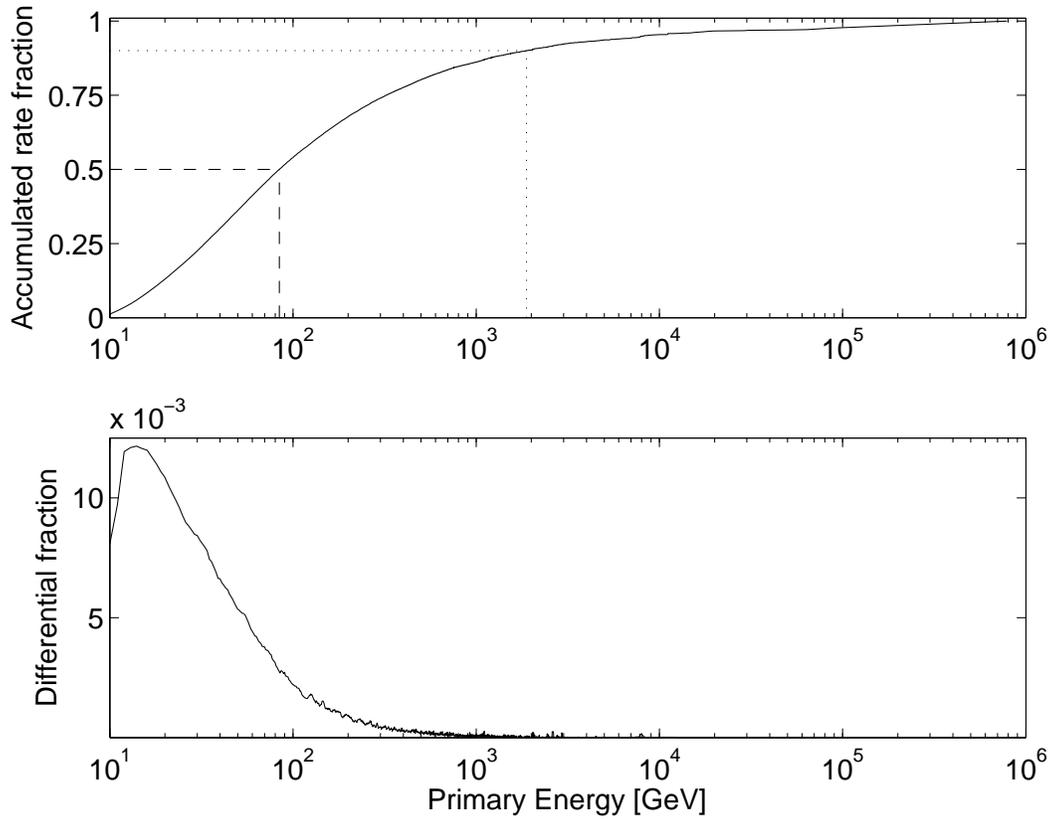}
\end{center}
\caption{Upper panel: Accumulated scaler counts in Period I as a function of
the energy of the primary cosmic ray assuming a stationary mean flux (see main
text).  Dashed and dotted lines show the $50\%$ and $90\%$ of counts,
respectively.  Lower panel: Differential response (derivative of the function
shown in the upper panel), obtained considering $dE=1$\,GeV (see main text).}
\label{figure_respuesta}
\end{figure}

The response of the detector is shown in Figure \ref{figure_respuesta}.  It was
computed from a set of low energy shower simulations using CORSIKA 6.980
\citep{Heck98} with the QGSJET-II model for high energy hadronic interactions
and GHEISHA low energy interaction routines.  The flux of primary particles at
the top of the atmosphere ($100$\,km of altitude) was simulated as a power law
in primary energy ($E_p$) with exponents obtained from the measured spectra in
the range  $10 \times Z_p < (E_p/\mathrm{GeV}) < 10^6$, and for $0^\mathrm{o}
\leq \theta_p \leq 88^\mathrm{o}$ in zenith angle \citep{grieder01}, for all
nuclei in the range $1 \leq Z_p \leq 26$ ($1\leq A_p \leq 56$).  The detector
response to the secondary particles was simulated using a simple simulator
developed within the Auger data analysis framework.  The detectors are
sensitive to charged particles from the shower of secondary particles at ground
level. This shower is essentially dominated by $\mu^\pm$ and e$^\pm$, as well
as to high energy photons that can be converted in $e^+$$e^-$ pairs before
being detected.

The simulated fraction of total counting rate produced by primaries with
kinetic energies lower than a specific energy (in the range
$10$\,GeV--$1$\,PeV) is shown in the upper panel of Figure
\ref{figure_respuesta}.  The dashed lines shows $50\%$ of the counts, which
corresponds to a median of $\sim 90$\,GeV. It also shows that primary particles
from $\sim 10$\,GeV (the geomagnetic rigidity cut off at Malarg\"ue is
$9.5$\,GV, see \cite{jinst}) up to $\sim 2$\,TeV produce $90\%$ of the counts
in the detector for Period I (dotted lines).  No significant differences in the
detector response have been observed for Period II.

The lower panel of Figure \ref{figure_respuesta} shows the detected count rate
fraction per unit of flux energy (i.e., the derivative of the curve shown in
the upper panel, which is frequently referred to the differential response
function), where a numerical derivative was computed using an increment of
$dE=1$\,GeV.  It shows that the largest sensitivity (per unit of log energy) of
the instrument corresponds to primary particles in the range
$10$\,GeV--$100$\,GeV.

\section{Comparison with Neutron Monitor Observations}\label{SCcomparison}

Neutron monitors (NMs) are standard detectors of cosmic rays at ground level.
In particular, the understanding of the modulation of CRs with energies larger
than $\sim$ 1 GeV arriving at Earth was significantly improved from
observations given by NMs \citep[see e.g., the review by][and references
therein]{Cane00}.  A sudden decrease of CRs observed at ground level (Forbush
decrease, Fd), with a gradual recovery lasting approximately one or two weeks,
occurs generally in coincidence with the passage of an interplanetary shock
or/and an ICME.

The amplitude of the daily variation of CRs observed by NMs generally increases
during the recovery phase of a Fd due to the presence of the anisotropy caused
by the transient interplanetary structure propagating beyond the Earth's orbit,
which produces a decrease of the cosmic ray flux arriving from that direction
\citep[e.g.,][and references therein]{lockwood71}.

\begin{figure}
\begin{center}
\includegraphics*[width=14cm,angle=0]{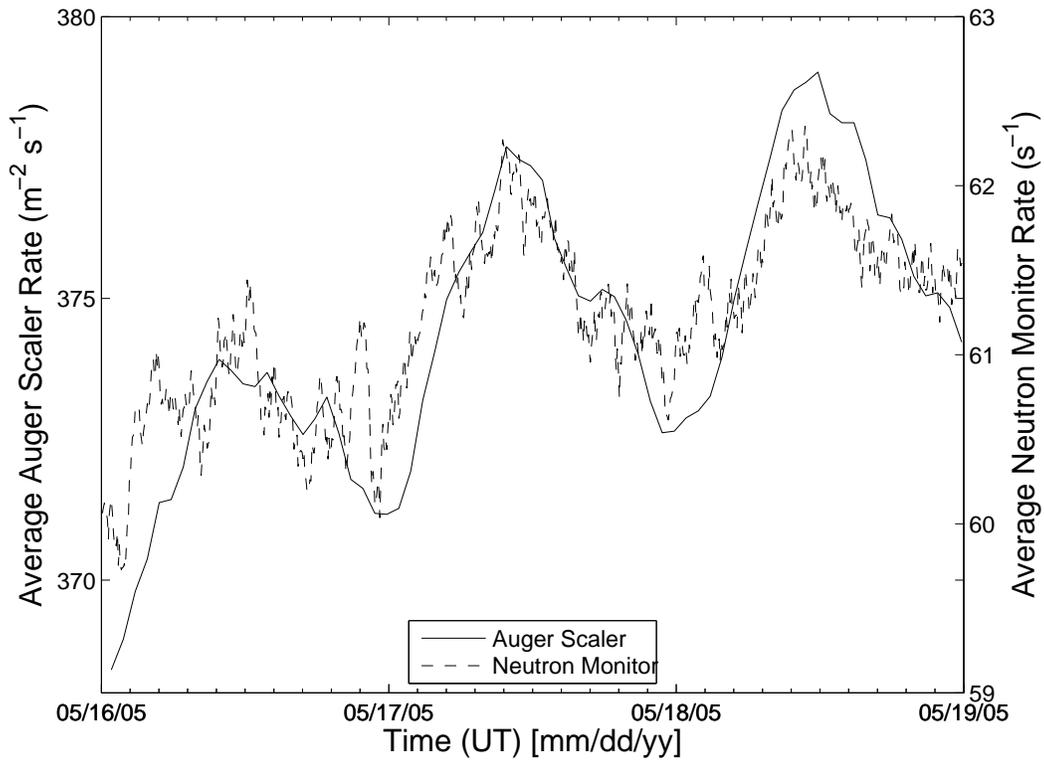}
\end{center}
\caption{Comparison of Auger scalers (solid line) with a neutron monitor
station (dashed line) for a period when the daily modulation is enhanced.}
\label{figure_compara_con_NMs}
\end{figure}

We present here the period after the Fd of May 2005 to make a comparison
between the daily variation observed with the scaler counter at the Auger
Observatory and a neutron monitor station.  In particular we chose the recovery
phase from May 16 to 19, 2005 (see Figure \ref{figure_compara_con_NMs}). 

The flux of secondary particles at ground level is significantly modulated by
the atmospheric pressure.  The scaler counts used and shown in this work are
pressure-corrected, as explained in \cite{jinst}.

In order to compare with observations at a similar location we chose neutron
monitor observations from Los Cerrillos Observatory 6NM64 (Chile), which is
located at $33.3^\mathrm{o}$\,S and $70.4^\mathrm{o}$\,W \citep{cordaro05},
only $\sim 250$\,km westward from the Auger Observatory.

The enhanced daily variation during the recovery phase of the analyzed Fd is
compared in Figure \ref{figure_compara_con_NMs} (for a comparison of the full
time range of this Fd, see \cite{jinst}), where a solid line represents the
average hourly Auger scaler counts, and the dashed line shows the counts
(hourly moving average over 5 minutes data points) observed at the NM in Los
Cerrillos Observatory.  An excellent agreement can be observed on the daily
variation.

We emphasize that the scalers from the Auger Observatory provide an absolute
flux of secondary particles with the high level of statistics provided by the
huge area of collection of particles. 

\section{Interplanetary cause of a Forbush decrease observed at the Auger
Observatory} \label{SCicme}

In this section we briefly analyze the transient interplanetary structure that
caused the Fd of May 2005 observed with the Auger scalers.

\begin{figure}
\includegraphics*[width=14cm,angle=0]{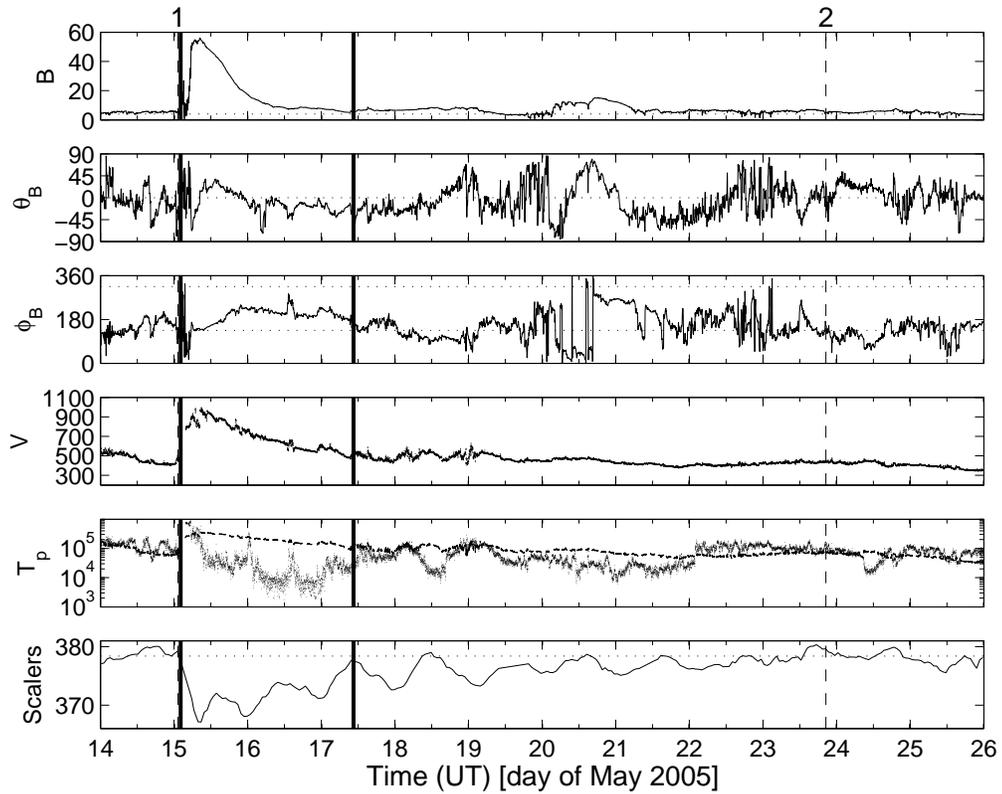}
\caption{From the upper panel: modulus [nT] of the IMF, latitude and longitude
of the IMF direction, bulk velocity [km/sec], and proton temperature [K]
(observed temperature marked with dots and expected temperature, $T_{ex}$,
marked with dashed line).  Lowest panel: Auger scaler counts per square meter
per second (time average of one hour).}
\label{figure_Fds_IP_parameters}
\end{figure}

Figure \ref{figure_Fds_IP_parameters} shows 'in situ' observations of the
interplanetary medium  obtained from the spacecraft ACE, which is located at
the Lagrangian point L1 in the solar wind near Earth, combined (lowest panel)
with the Auger scaler rate for the time range around the Forbush event of 15
May 2005.

The solar wind data were obtained from the Magnetic Fields Experiment (MAG)
\citep{Smith98} and the plasma data from the Solar Wind Electron Proton Alpha
Monitor (SWEPAM) \citep{McComas98}.  In the upper panel the figure shows the
modulus ($B$) and orientation (latitude $\theta_B$ and longitude $\phi_B$) of
the interplanetary magnetic field (IMF) direction in the Geocentric Solar
Ecliptic system, GSE.  Then, it shows plasma conditions of the interplanetary
medium (bulk velocity $V$ and proton temperature $T_p$), and the scaler count
data.

Data in Figure \ref{figure_Fds_IP_parameters} clearly show the presence of an
interplanetary mass ejection between May 15 and May 17.  Inside this time range
a large scale smoothly varying magnetic field orientation of high intensity can
be observed, typical of a subset of ICMEs called magnetic clouds
\citep{Burlaga81}.  The bulk velocity for this ICME is in agreement with an
expanding ICME, traveling faster in the front than in the back, presenting an
almost linear bulk velocity profile, as expected for typical ICMEs
\citep{Demoulin08}.  The observed proton temperature is lower than the expected
temperature ($T_{ex}$) for a typical solar wind at a given observed bulk
velocity \citep{lopez87,demoulin09}, in agreement with typical observations of
an ICME \citep{richardson95}.

The arrival of the ICME was previously identified at 02:11 UT on 15 May
\citep{dasso09}.  The temporal length of this ICME corresponds to a very
extended region, $\sim 0.93$\,AU, one of the largest ICMEs ever observed.  The
magnetic field in this ICME is huge, reaching almost $60$\,nT, one of the
largest values observed in the interplanetary medium at $1$\,AU.  At 10:30 UT
of 17 of May, the magnetic field recovers its background value of $\sim 5$\,nT
and typical solar wind conditions.  Thick vertical solid lines in Figure
\ref{figure_Fds_IP_parameters} show the boundaries of this huge ICME, which was
formed by two flux ropes \citep{dasso09} and their strongly perturbed
environment.

Some simple models for describing arrival of CRs at Earth are based on
diffusive obstacles (or barriers \citep{wibberenz00}) that obstruct the
transport of low energy cosmic rays.  These barriers result from perturbed
solar wind associated with ICMEs, and they have a diffusion coefficient
depending on the interplanetary magnetic field intensity.  It has been proposed
that these diffusive barriers are the main origin of the weakened flux of CRs
observed at Earth during a Fd \citep[e.g.,][and references therein]{Cane00}.

Scaler counts (last panel of Figure \ref{figure_Fds_IP_parameters}) show the structure 
of this Fd, which starts at 01:18 UT on 15 May 2005 and 
ends at 20:30 UT on 23 May 2005, as has been reported in \cite{jinst}.
Start and end times of the Fd are marked with ticks '1' and '2' and also 
as vertical dashed lines in Figure \ref{figure_Fds_IP_parameters}.
The start time (1) is in agreement with the arrival of the interplanetary 
shock driven by this huge ICME.

The magnetic field ($B$) in ICMEs plays an important role in the transport of
CRs.  Its typical decay for solar distances ($D$) between $0.3$\,AU and $5$\,AU
has been quantified as $B(D) \sim B_0 (D/D_0)^{-1.5}$ \citep{Wang05}, with
$B_0$ the reference magnetic field at a reference distance $D_0$.  This decay
is a direct consequence of the increase of the ICME size with $D$ and the
conservation of the magnetic flux across a material surface. The increase of
the ICME size in the direction perpendicular to its main axis is mainly driven
by the decrease of the environment solar wind pressure
\citep{Demoulin09b,Gulisano10}.  Its typical size ($S$) along the radial
direction from the Sun evolves as $S(D) \sim S_0 (D/D_0)^{0.6}$ \citep{Wang05}.

The ICME observed during May 15-18, traveled with a mean bulk speed of $\sim$
700 km/sec.  According to the scalers (shown in the last panel at the bottom of
Figure \ref{figure_Fds_IP_parameters}) the Forbush decrease finished around
23-24 May (see also \cite{jinst}). Thus, assuming that the ICME shown in Figure
\ref{figure_Fds_IP_parameters}  will evolve as a typical one, at the moment of
the end of the decay phase of the Fd (when it finished its effect on the
weakening of CRs flux at Earth), it could have been at a distance $D \sim 4.2$
AU from the Sun, with a radial size of $S \sim 2.2$ AU, and a maximum magnetic
field of the order of $B \sim 7$ nT, while a typical solar wind magnetic field
at this solar distances is lower than $1$\,nT.
  
\section{Summary and Conclusions} \label{SCconclusion}

The heliosphere provides a natural laboratory where it is possible to study a
huge number of physical processes, with the help of fleets of spacecraft (which
can observe remote and 'in situ' properties of the plasma in the interplanetary
medium) and with the increasing number of new observatories at ground level
(which can observe electromagnetic radiation from the local cosmos and cosmic
ray fluxes). The combination of results from these new generation observatories
of increasing quality will allow us to improve our heliospheric models and
achieve significant progress in our understanding of space physics during the
next years.

The Pierre Auger Observatory provides scaler data, which measure the flux of
low energy cosmic rays with a very high statistical significance from its huge
total collecting area of more than $16\,000$\,m$^2$, which provides a
consequent $\sim 10^8$\,counts per minute and a statistical accuracy well below
$1\permil$ with the scalers \citep{jinst}.

These observations correspond to a wide primary energy range of three order of
magnitudes, from $\sim 10$\,GeV to $\sim 10$\,TeV (see Figure
\ref{figure_respuesta}).  Primary particles in the range of $\sim 10$\,GeV and
$\sim 1$\,TeV give the maximal ($\sim 90\%$ of the scaler rate fraction)
contribution to the signal, with a median primary energy of $\sim 90$\,GeV.

Thus, scalers provide data-sets complementary to networks of neutron monitors
(which have lower effective/median energy) and nearer to networks of muon
detectors (effective/median energy $>50$ GeV), with scalers having a
significantly higher statistical accuracy.

From a detailed comparison of the Auger scalers with a neutron monitor station
near Malarg\"ue, during a period when the daily modulation is enhanced at the
decay phase of a Forbush decrease, we observe an excellent agreement on the
phase of this modulation.

The combination of ground and space observations has allowed us to analyze some
properties of the ICME and the interplanetary shock that caused one of the
Forbush decreases observed at Auger.  It was a very large ICME at distances of
$1$\,AU from the Sun (almost $\sim 1$\,AU of size in the Sun-Earth direction)
with one of the highest magnetic fields ever observed.

The 15 minutes time averaged Auger scalers, averaged over the whole surface
detector array, are publicly available and can be downloaded from the Pierre
Auger Observatory Public Event Display site {\small [http://www.auger.org]}.
This web interface permits visualization and downloading of the data.

\section{Acknowledgments}
\label{SCacknowlefges}

We are grateful to Prof. Enrique G. Cordaro and the Observatorio de Radiaci\'on
C\'osmica of the Universidad de Chile, who kindly provided the neutron monitor
data of Los Cerrillos Observatory, and to the ACE-SWEPAM and ACE-MAG teams for
the data used for this work.  We thank the reviewers for very useful comments,
which have permitted to improve the quality of the manuscript.  S. Dasso
acknowledges support from the Abdus Salam International Centre for Theoretical
Physics (ICTP), as provided in the frame of his regular associateship.

 \end{document}